\documentclass[aps,prb,twocolumn,superscriptaddress,showpacs]{revtex4}
\usepackage{graphicx}
\begin{document}

\title{First-order transition in the itinerant ferromagnet CoS$_{1.9}$Se$_{0.1}$}

\author{T. J. Sato}
    \email[]{tjsato@nist.gov}
    \altaffiliation{On leave from the National Institute for Materials Science, Tsukuba 305-0047, Japan.}
\affiliation{NIST Center for Neutron Research, National Institute of Standards and Technology, Gaithersburg, MD 20899-8562}
\author{J. W. Lynn}
\affiliation{NIST Center for Neutron Research, National Institute of Standards and Technology, Gaithersburg, MD 20899-8562}
\affiliation{Department of Physics, University of Maryland, College Park, MD 20742}
\author{Y. S. Hor}
\author{S. -W. Cheong}
\affiliation{Department of Physics and Astronomy, Rutgers University, Piscataway, NJ 08854}

\date{\today}

\begin{abstract}
Undoped CoS$_2$ is an isotropic itinerant ferromagnet with a continuous or nearly continuous phase transition at $T_{\rm C} = 122$~K.
In the doped CoS$_{1.9}$Se$_{0.1}$ system, the Curie temperature is lowered to $T_{\rm C} = 90$~K, and the transition becomes clearly first order in nature.
In particular we find a discontinuous evolution of the spin dynamics as well as strong time relaxation in the ferromagnetic Bragg intensity and small angle neutron scattering in vicinity of the ferromagnetic transition.
In the ordered state the long-wavelength spin excitations were found to be conventional ferromagnetic spin-waves with negligible spin-wave gap ($ < 0.04$~meV), indicating that this system is also an excellent isotropic (soft) ferromagnet.
In a wide temperature range up to $0.9T_{\rm C}$, the spin-wave stiffness $D(T)$ follows the prediction of the two-magnon interaction theory, $D(T) = D(0)(1 - AT^{5/2})$, with $D(0) = 131.7 \pm 2.8$~meV-\AA$^{2}$.
The stiffness, however, does not collapse as $T \rightarrow T_{\rm C}$ from below.
Instead a quasielastic central peak abruptly develops in the excitation spectrum, quite similar to results found in the colossal magnetoresistance oxides such as (La-Ca)MnO$_3$.
\end{abstract}


\maketitle

\section{Introduction}
The pyrite-type transition-metal disulfides $M$S$_2$ ($M =$ Fe, Co, Ni, Cu and Zn) exhibit a wide variety of magnetic and electronic ground states~\cite{jar68,ada69,gau75,mat00}, ranging from a paramagnetic or antiferromagnetic semiconductor in FeS$_2$ or NiS$_2$ to a superconductor in CuS$_2$.
Among them, the metallic ferromagnet CoS$_2$ has attracted revived interest recently because of its unique electronic structure~\cite{kwo00,shi01}.
Owing to a strong crystal field, the Co $3d$ bands split into lower $t_{2g}$ and higher $e_{g}$ bands.
The Fermi level crosses the $e_g$ band, resulting in fully-occupied $t_{2g}$ and quarter-filled $e_{g}$ bands.
In an ionic picture, this electronic configuration corresponds to the low-spin state of the Co$^{2+}$ $(3d)^7$ electrons~\cite{bit68}, {\it i.e.}, $t_{2g}^6e_g^1$ ($S = 1/2$).
The $e_g$ electrons, which are the only source of magnetism in CoS$_2$, are also responsible for electron conduction through hybridization with antibonding $p^{*}$ states arising from the sulfur pairs~\cite{fol87}.
Hence mutual interplay of electron transport and magnetism can be expected.

CoS$_2$ exhibits ferromagnetic ordering at the Curie temperature $T_{\rm C} = 122$~K.
The saturation moment is about $0.86~\mu_{\rm B}$ per Co atom~\cite{jar68,ada69}, which is close to the full polarization of one $e_g$ electron ($1~\mu_{\rm B}$).
Indeed, recent optical- and photoemission-spectroscopies indicate a nearly (or even fully) spin polarized $e_g$ band~\cite{yam99,tak01}, suggesting CoS$_2$ as a candidate for the so-called half-metallic ferromagnet~\cite{pri95,par98}.
The spin dynamics of CoS$_2$ has been explored by two groups using neutron scattering technique.
Iizumi {\it et al.} showed that low-energy spin excitations are conventional ferromagnetic spin-waves~\cite{iiz75}.
The spin-wave energy gap is negligibly small, indicating that the system is isotropic, as expected for a $S = 1/2$ system with one $e_g$ electron.
In addition, the existence of Stoner excitations has been inferred at higher energies ($E > 8$~meV) by Hiraka {\it et al.}~\cite{hir97}, evidencing the itinerant nature of the ferromagnetism in CoS$_2$.

It is expected that the hybridization between the $e_g$ and $p^{*}$ states is stronger in CoSe$_2$ than in CoS$_2$~\cite{got97,yam98}.
Therefore, partial substitution (doping) of S by Se makes the $e_g$ band wider (or increases the transfer integral $t$ in the ionic picture), resulting in a decrease of the density of states (DOS) at the Fermi level.
To date, several experiments have been performed to investigate the Se-doping effect~\cite{ada69,ada70,pan79,hir96,got97}.
They all report strong suppression of the ferromagnetic transition in the Se-doped Co(S$_{1-x}$Se$_x$)$_2$ compounds; $T_{\rm C}$ decreases rapidly as $x$ increases, and the ferromagnetic phase disappears ($T_{\rm C} = 0$) at $x = 0.12$.
Interestingly, we have found that even for very small Se-concentrations ({\it e.g.} $x = 0.05$) the ferromagnetic transition becomes first-order.
Field-induced metamagnetic transitions were also reported in the Se-doped compounds~\cite{got97}.

The earlier studies are mainly aimed at investigating the magnetic phase diagram using magnetization measurements.
The effect of Se doping on the spin dynamics is totally unknown to date.
Therefore in the present study we have performed neutron scattering experiments on the Se-doped sample Co(S$_{1-x}$Se$_x$)$_2$ with $x = 0.05$.
We have confirmed the first-order nature of the ferromagnetic transition at $T_{\rm C} = 90$~K using elastic neutron scattering experiments. 
Detailed measurements of the magnetic excitation spectrum were carried out in a wide temperature range of $T \leq 92.5$~K and an energy range of $E \leq 1.5$~meV.
A prominent feature is the abrupt development of a quasielastic central peak, accompanied by a decrease of conventional spin-wave-peak intensity, observed in the vicinity just below $T_{\rm C}$.
This suggests that the ferromagnetic phase abruptly transforms to paramagnetic fluctuations, instead of the continuous thermal population of spin-waves that occurs in a second-order transition.
This behavior is quite similar to ferromagnetic transitions in colossal magnetoresistance manganites.

\section{Experimental}
A polycrystalline sample of CoS$_{1.9}$Se$_{0.1}$ was prepared by using a solid-state reaction technique. 
The starting materials, Co, S, and Se powders, were mixed in the calculated ratio.
After mixing, they were ground, pressed into a pellet, sealed in a vacuum quartz tube, and heated to 650$^{\circ}$~C for a period of 168 hours. 
Subsequently, the resultant material was reground, pressed into pellet and was heated for 150 hours in a sealed vacuum quartz tube at 700$^{\circ}$~C and 750$^{\circ}$~C for the second- and third-sintering, respectively. 
X-ray diffraction confirmed that the sample is the single-phase pyrite structure.
Rsistivity measurements on the polycrystalline sample were taken using a conventional four-probe technique with an excitation current of 1~mA.

All neutron scattering experiments were performed at National Institute of Standards and Technology Center for Neutron Research. 
A powder sample of CoS$_{1.9}$Se$_{0.1}$ (about 12~g) was loaded in an Al cell with the sample-space size of 32~mm (height) $\times$ 28~mm (width) $\times$ 6~mm (thickness).
The small thickness is to reduce the effects of the strong neutron absorption of Co.
The sample cell was then attached to a closed cycle refrigerator or a $^4$He cryostat.
For small angle neutron scattering (SANS) and forward direction inelastic scattering experiments, air and vacuum chamber walls of cryostats in the beam path can be possible sources for the background.
Therefore, we used a large diameter vacuum chamber with thin single-crystal Si windows to reduce the background.

Elastic experiments were performed using the BT-7 triple-axis spectrometer.
A double-crystal monochromator with pyrolytic graphite (PG) 002 reflections and a single-crystal PG 002 analyzer were used to fix the incident and outgoing neutrons to 13.46~meV.
A PG filter was placed before the monochromator to reduce higher wavelength contaminations.
The effective horizontal collimation before the sample was 30', and 40' after the sample for the elastic experiments.
Small angle neutron scattering data were obtained on the NG-7 30~m SANS spectrometer, where the incident neutron wavelength was chosen as 10 or 14~\AA, and the detector position was 8~m.

Inelastic experiments were carried out using the BT-2 and BT-9 triple-axis spectrometers. 
The PG 002 reflections were used for both the monochromator and analyzer, and higher order neutrons were eliminated by the PG filter.
For isotropic ferromagnets, long-wavelength spin-wave excitations can be observed by using the powder sample in the forward direction, {\it i.e.}, around the 000 reciprocal lattice point, without loss of generality~\cite{lyn95}.
For this type of measurement, tight horizontal collimations of 20'-10'-10'-20' (BT-2) or 10'-10'-10'-10' (BT-9) were employed to reduce the undesirable background from the direct beam and achieve adequate instrumental resolution.
The inelastic magnetic scattering intensity is strongly suppressed at the lowest temperature, compared to that at around $T_{\rm C} = 90$~K, due to reduction of thermal population factor of spin-wave excitations in the present energy range ($E \leq 1.5$~meV).
Thus, the lowest temperature data, which were typically taken below 10~K, can be an estimate of non-magnetic scattering.
The non-magnetic scattering is then subtracted from the higher temperature data, and the remaining intensity will be shown as inelastic magnetic scattering spectra in the following.
A triple-axis spectrometer generally has quite relaxed vertical collimations to increase the counting rate.
This, however, gives a large resolution volume for the vertical direction, and for these small-$q$ spin-wave measurements, the instrumental resolution must be taken into account to deduce the spin-wave peak energy from the experimentally obtained spectra.
To accomplish this, we performed resolution-convoluted least-squares fits using a model scattering function $S(Q, \hbar\omega)$.
As a model $S(Q, \hbar\omega)$ we assumed the following function comprising the quasielastic as well as the inelastic Lorentzian peaks:
\begin{eqnarray}\label{fittingfunction}
	S(Q, \hbar\omega) &\propto& 
	\hbar\omega[1 + n(\hbar\omega)]
	\left[ \frac{I_{\rm q} \Gamma_{\rm q}}
	{\Gamma_{\rm q}^2 + \hbar\omega^2} \right. \nonumber \\
	 &+& \left. 
	\frac{I\Gamma}
	{\Gamma^2 + (\hbar\omega + E_{q})^2}+
	\frac{I\Gamma}
	{\Gamma^2 + (\hbar\omega - E_{q})^2}\right],
\end{eqnarray}
where $[1 + n(\hbar\omega)] (= [1 -\exp(-\hbar\omega/k_{\rm B}T)]^{-1})$ is the Bose temperature factor.
For $E_{q}$, we used the dispersion relation of conventional ferromagnetic spin-waves without an anisotropy gap, {\it i.e.}, $E_{q} = D(T)q^2$, where $D(T)$ is the spin-wave stiffness parameter.

\section{Experimental results}

\subsection{Elastic experiments}
\begin{figure}
 \includegraphics[scale=0.25, angle=-90]{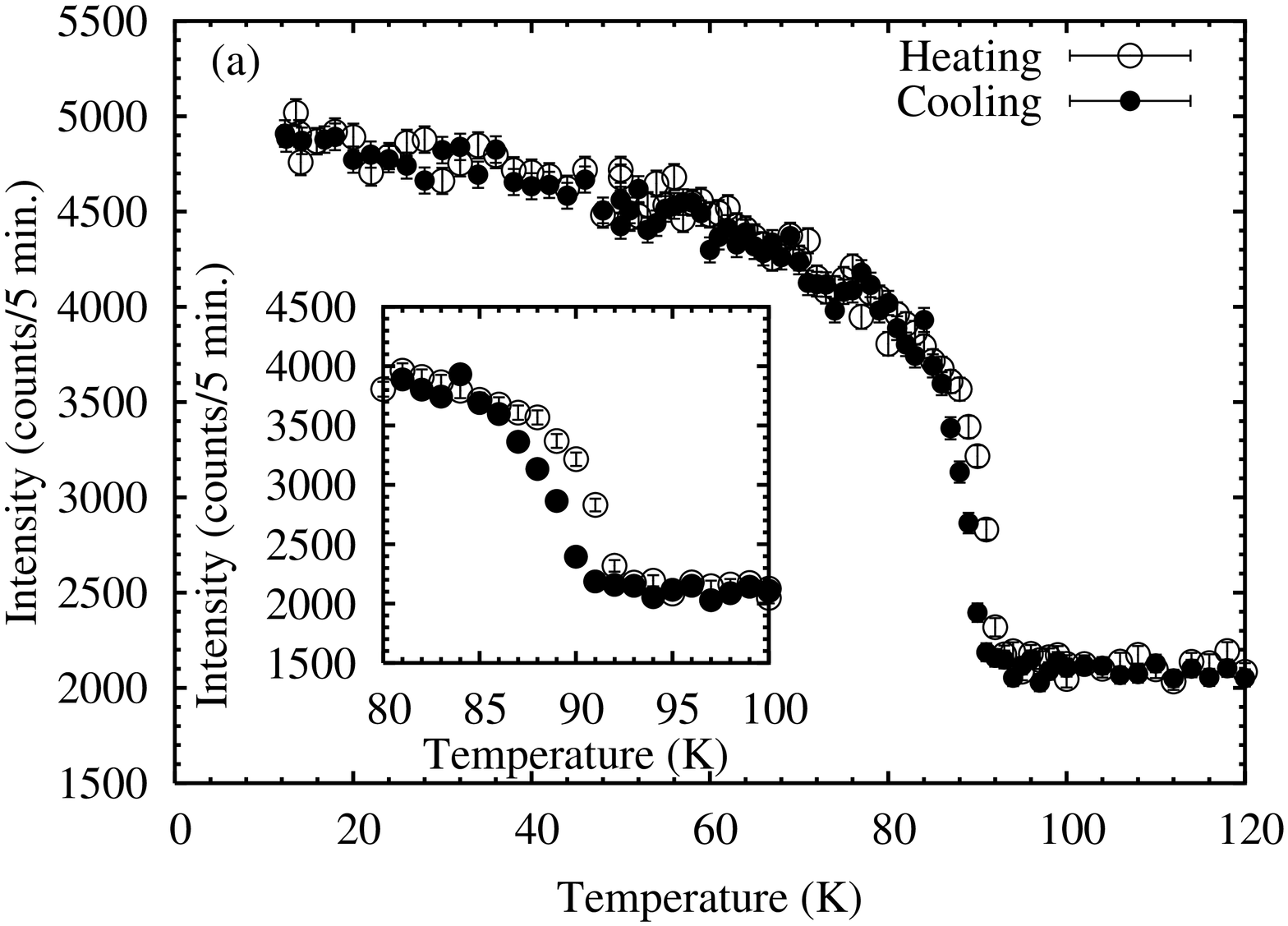}%

 \includegraphics[scale=0.63, angle=0]{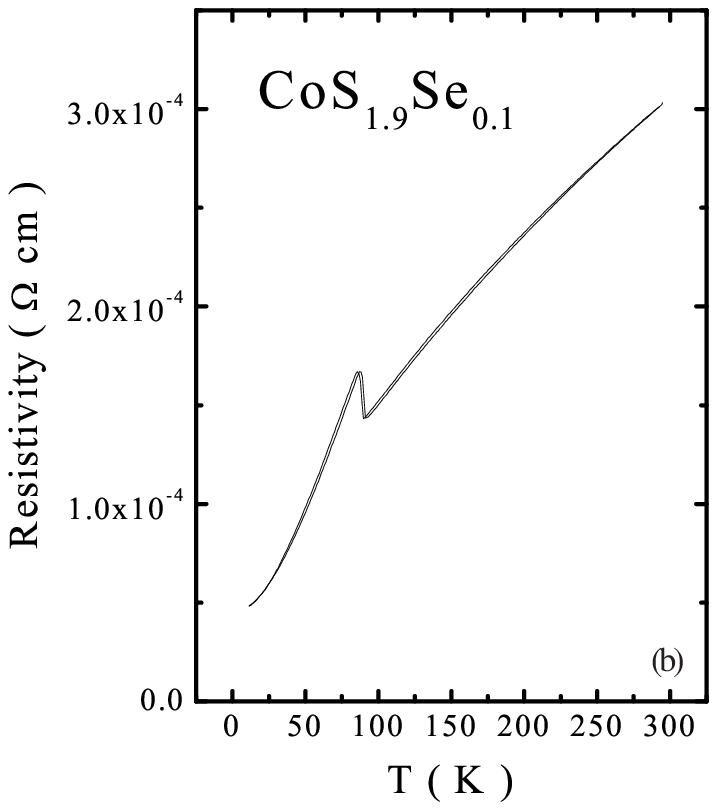}%

 \includegraphics[scale=0.25, angle=-90]{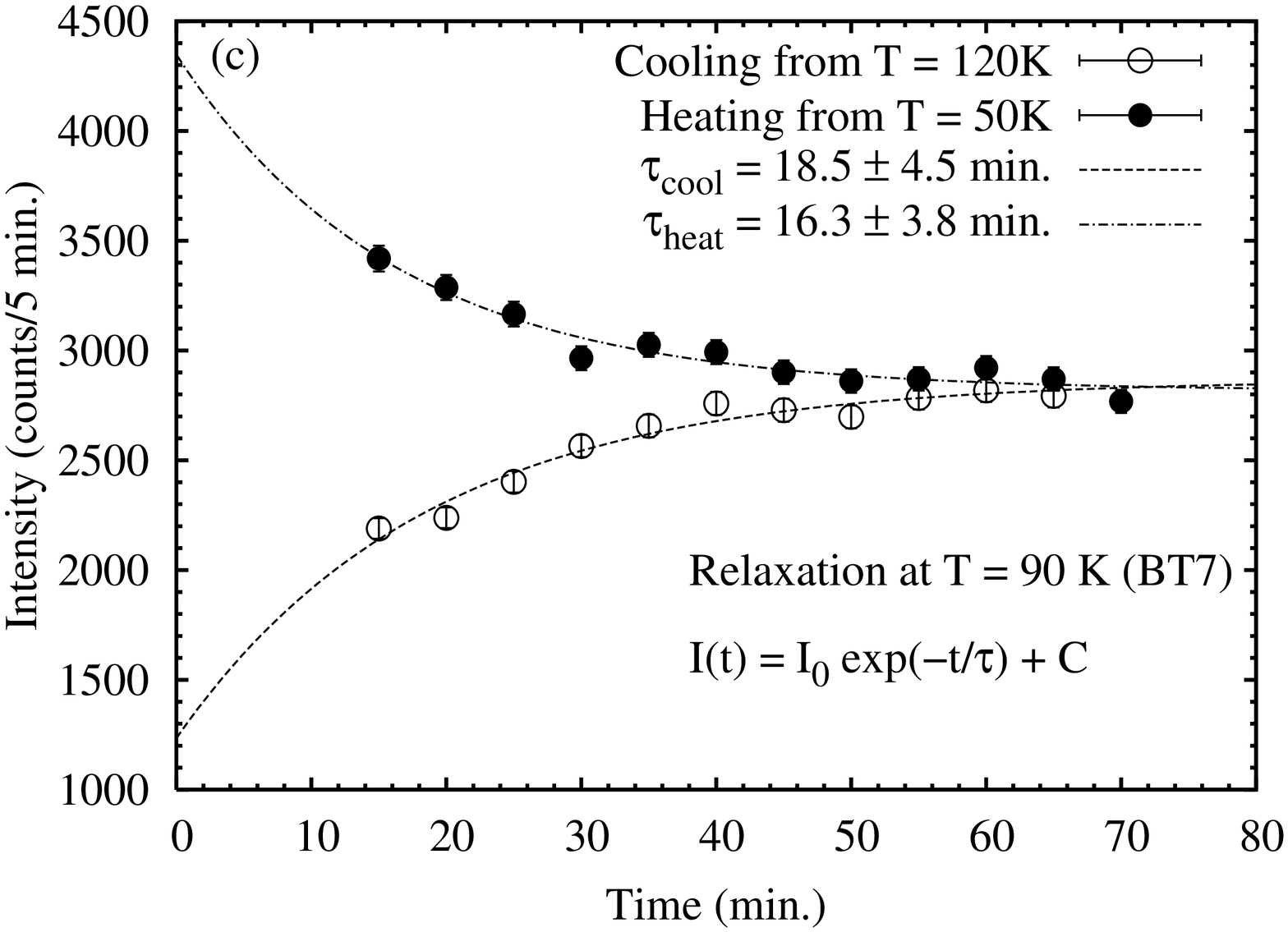}%
 \caption
{(a) Temperature dependence of the 111 intensity. 
	Open and filled circles stand for the heating and cooling runs, respectively.  
	At each temperature, we waited two minutes for temperature stabilization before counting for five minutes.
	Inset: magnified plot around the transition temperature $T_{\rm C} = 90$~K.
	Hysteresis effects can be clearly seen.
	(b) Temperature dependence of the resistivity.
	Two solid lines stand for cooling and heating results measured with the rates of 3~K/min and 6~K/min, respectively.
	The magnetic phase transition can be observed as a sharp anomaly at $T_{\rm C} = 90 $~K.
	Some thermal irreversibility was observed around $T_{\rm C}$ in these measurements.
	(c) Time relaxation of the 111 reflection at $T = 90$~K $= T_{\rm C}$.
	Open circles are for the relaxation after cooling from the paramagnetic phase ($T = 120$~K), whereas filled circles are for the relaxation after heating from the ferromagnetic phase ($T = 50$~K).
	Dashed and dash-dotted lines are fits to the exponential decay $I(t) = I_0\exp(-t/\tau) + C$.}
\end{figure}

We determined the ferromagnetic transition temperature of the $x = 0.05$ sample by measuring the temperature dependence of the magnetic Bragg intensity.
In ferromagnets, the magnetic reflections appear at the same $Q$ positions as the nuclear Bragg reflections, and for unpolarized neutrons, the intensities add.
We used the 111 reflection for this measurement, where the nuclear intensity is weak and thus the relative magnetic signal should be optimal.
For a first-order transition it is crucial to perform temperature scans in a well-controlled manner; in particular, heating and cooling scans should be measured using the identical scanning rate.
Thus, for both scans the temperature step was fixed to 1~K, and at each temperature we waited two minutes for temperature stabilization before counting for five minutes.
The resulting temperature dependence of the 111 intensity is shown in Fig.~1(a).
At higher temperatures, there exists a temperature independent intensity of about 2000~cts/5~min, which is attributed to the nuclear 111 reflection.
As the temperature is decreased, the intensity abruptly increases at $T \sim 90$~K, indicating establishment of the ferromagnetically ordered phase.
The sharpness of the transition suggests a negligible distribution of the transition temperatures, confirming chemical homogeneity.
On the other hand for the heating runs, the magnetic intensity remains finite up to the slightly higher temperature $T \sim 92$~K.
This thermal hysteresis is clearly seen in the magnified plot around the ferromagnetic transition shown in the inset of Fig.~1(a).
By taking the center of the hysteresis loop, we determine the Curie temperature as $T_{\rm C} = 90(1)$~K for the $x = 0.05$ sample.
The ferromagnetic transition was also readily observed in resistivity measurements. 
Fig.~1(b) shows the temperature dependence of the resistivity, in which a sharp anomaly is seen at $T_{\rm C} = 90$~K.
A similar anomaly was observed in the pure CoS$_2$ system at $T \simeq 120$~K, and was attributed to an electronic structure change at the Fermi level due to the ferromagnetic transition~\cite{yom79}.

Next, we investigate the time scale for thermal equilibration.
For this purpose, the time relaxation of the 111 intensity was measured at $T_{\rm C}$ after heating from the ferromagnetic phase ($T = 50$~K) or after cooling from the paramagnetic phase ($T = 120$~K).
Fig.~1(c) shows the resulting time relaxations.
The origin of the horizontal axis, {\it i.e.} $t = 0$, was chosen to be the time when the sample temperature became within $T_{\rm C} \pm3$~K.
The scattering intensity after cooling (heating) exhibits a monotonic increase (decrease), which can be well approximated by an exponential decay $I(t) = I_0 \exp(-t/\tau) + C$, shown by the dashed and dash-dotted lines, with the characteristic time scale $\tau \sim 17$~min.
This strong relaxation effect indicates the first-order nature of the ferroamgnetic transition.
Both the relaxation curves merge asymptotically after about one hour.
Therefore no genuine hysteresis, {\it i.e.} true hysteresis in the asymptotic region, is observed.
Similar behavior has been found in the colossal magnetoresistive manganites~\cite{ada03}.
Thermal relaxation behavior was also observed in the resistivity measurements, but no attempt was made to separate the intrinsic relaxation effects from the thermal lag of the apparatus.

\begin{figure}
 \includegraphics[scale=0.3, angle=-90]{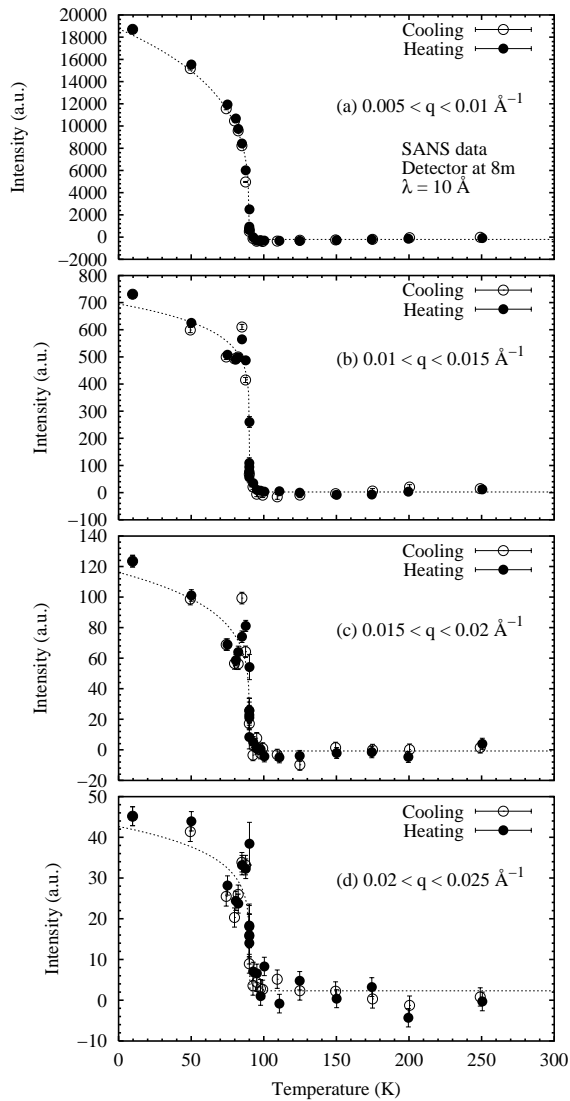}
 \caption{Temperature dependence of the energy-integrated small angle magnetic scattering.
	$q$-integration was made for (a) $0.005 < q < 0.01$~\AA$^{-1}$, (b) $0.01 < q < 0.015$~\AA$^{-1}$, (c) $0.015 < q < 0.02$~\AA$^{-1}$, and (d) $0.02 < q < 0.025$~\AA$^{-1}$.
	The non-magnetic background was subtracted using the paramagnetic data obtained at $300~$K.
	Open and closed circles stand for the cooling and heating runs, whereas dotted lines are guides for the eyes.
	}
\end{figure}

\begin{figure}
 \includegraphics[scale=0.3, angle=-90]{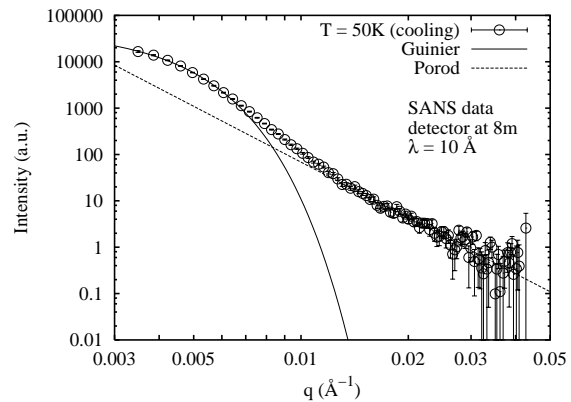}
 \caption{Small angle scattering intensity versus $q$ measured at $T = 50$~K $< T_{\rm C}$.
	The solid line stands for the fit to the Guinier law, $I(q) \propto \exp(-R^2q^2/3)$, whereas the dashed line represents the fit to the Porod law $I(q) \propto q^{-4}$.
	Non-magnetic component was subtracted using the paramagnetic $T = 300$~K data.
	}
\end{figure}

The discontinuous nature of the ferromagnetic transition was prominently observed in the SANS data.
Fig.~2 shows the temperature dependence of the SANS intensities integrated in the four $q$-ranges, $0.005 < q < 0.01$, $0.01 < q < 0.015$, $0.015 < q < 0.02$ and $0.02 < q < 0.025$~\AA$^{-1}$.
These figures only show the magnetic scattering component; the non-magnetic background was subtracted by using the paramagnetic $T = 300$~K data.
Irrespective of the $q$-ranges, the scattering intensity abruptly develops below $T_{\rm C} = 90$~K.
We also confirmed the strong relaxation effect of the SANS intensity at $T_{\rm C}$, similar to the ferromagnetic order parameter.
Fig.~3 shows the $q$-dependence of the SANS intensity in the ferromagnetic phase measured at $T = 50$~K.
The $q$-dependence is well reproduced by the Porod form $I \propto q^{-4}$ at higher $q$ ($q > 0.015$~\AA$^{-1}$), whereas it obeys the Guinier law $I(q) \propto \exp(-R^2q^2/3)$ in the low-$q$ region ($q < 0.006$~\AA$^{-1}$).
These $q$ dependencies indicate that the scattering is due to ferromagnetic domains with the length scale $R \sim 500$~\AA.

An outstanding feature of the SANS data is an absence of critical scattering; as seen in Fig.~2, the SANS intensity above $T_{\rm C}$ is negligibly small in the present $q$-range ($q < 0.04$~\AA$^{-1}$), suggesting that the energy integrated magnetic scattering intensity has a negligible temperature dependence for $T > T_{\rm C}$.
This is in striking contrast to conventional ferromagnets with a second-order transition, where the strongly temperature-dependent critical scattering of the Ornstein-Zernike form, $I(q) \propto 1/(q^2 + \kappa^2)$, can be observed in the forward direction near $T_{\rm C}$ (see experiments on Tl$_2$Mn$_2$O$_7$ as an example~\cite{lyn98}.)
The absence of the critical scattering is additional evidence of the first-order nature of the ferromagnetic transition in CoS$_{1.9}$Se$_{0.1}$.

\subsection{Inelastic experiments}
\begin{figure}
 \includegraphics[scale=0.5, angle=-90]{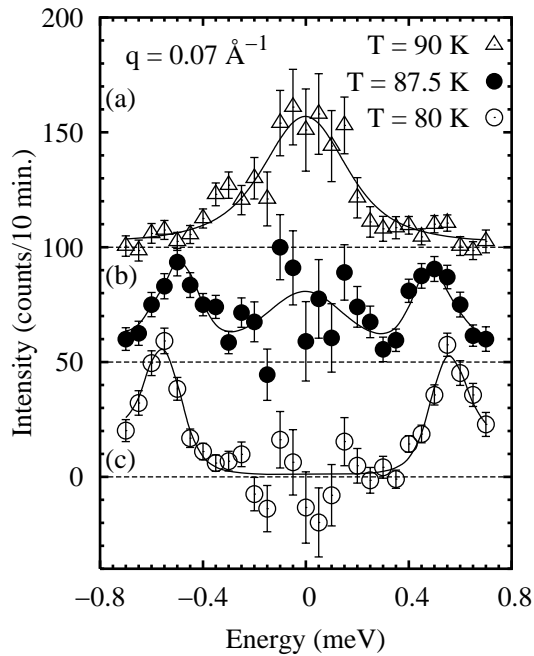}%
 \caption{Inelastic scattering spectra at temperatures of 80~K, 87.5~K and 90~K and $q = 0.07$~\AA$^{-1}$.
	Non-magnetic scattering was subtracted using the lowest temperature spectrum as described in the text.
	The data at $87.5$~K and $90$~K are shifted upward by 50 and 100 counts, respectively, to increase the visibility.
	The solid curves are fits to Eq.~(1) convoluted with the instrumental resolution function.
	The central component coexists with the spin-wave peaks at $T = 87.5$~K.
	}
\end{figure}

\begin{figure}
 \includegraphics[scale=0.3, angle=-90]{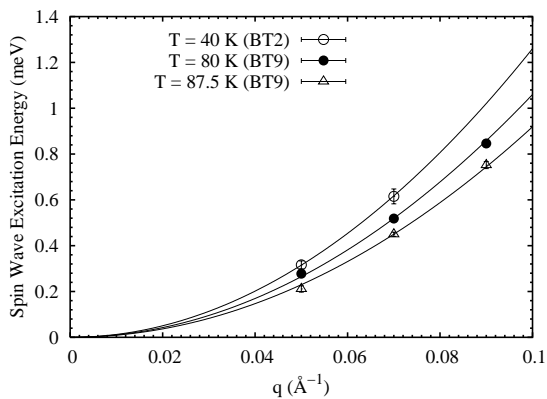}%
 \caption{Spin-wave excitation energy versus $q$ at the three temperatures $T = 40$, 80 and 87.5~K.
	The solid lines represent a fit to the quadratic dispersion relation $E_q = D(T)q^2$.
	The absence of a significant gap indicates that CoS$_{1.9}$Se$_{0.1}$ is an excellent isotropic ferromagnet.}
\end{figure}

We now turn to the inelastic scattering measurements to explore the spin fluctuation spectrum in Co(S$_{1-x}$Se$_x$)$_2$.
A number of constant-$q$ scans were performed at a series of wave-vector transfers ($q$'s) in the temperature range $T \leq 92.5$~K.
Shown in Fig.~4 are representative inelastic spectra taken at the three temperatures $T = 80$, 87.5 and 90~K with $q = 0.07$~\AA$^{-1}$.
At $T = 80$~K $<< T_{\rm C}$, well-defined spin-wave peaks can be seen in the neutron-energy gain ($E < 0$) and loss ($E > 0$) sides at $E \sim \pm 0.6$~meV (Fig.~4(c)).
The two peaks show almost identical intensity, whereas no central component can be seen around $E=0$~meV.
These results confirm the validity of our background subtraction procedure.
The solid curve is the result of a resolution-convoluted least-squares fit to the model cross-section Eq.~(1).
The excitation energy was obtained from the fitting as $E_{q = 0.07} = 0.52$~meV at $T = 80$~K, which is slightly lower than the apparent peak position.
This reduction in energy represents the correction for the resolution.
The inelastic spectrum at $T = 87.5~{\rm K} (< T_{\rm C})$ is shown in Fig.~4(b).
As the temperature becomes closer to $T_{\rm C}$, the spin-wave peaks shift toward the elastic position, exhibiting spin-wave renormalization.
Simultaneously, the peak width becomes broader, indicating shorter lifetimes of spin-wave excitations.
These features are in qualitative agreement with the expected temperature dependence of conventional spin-waves.
In addition, there appears a quasielastic central peak at $T = 87.5$~K.
It should be noted that the spectrum was measured a few hours after the temperature was set to 87.5~K, and thus this coexistence of the quasielastic and spin-wave peaks cannot be a transient effect associated with the first-order transition.
Instead, appearance of the quasielastic component suggests that there intrinsically exist two kinds of spin fluctuations in the vicinity of $T_{\rm C}$.
By further increasing the temperature, the quasielastic component dominates the magnetic fluctuations, as exemplified by the spectrum taken at $T = T_{\rm C} = 90$~K shown in Fig.~4(a).

Let us first discuss the spin-wave excitations observed below $T_{\rm C}$.
Fig.~5 shows the spin-wave dispersion relations measured at the three temperatures $T = 40$, 80 and 87.5~K.
As evidenced in the figure, the excitation energy obeys the usual quadratic dispersion relation $E_{q} = \Delta + D(T)q^2$.
We further found that the gap $\Delta$ is too small to be determined in the present thermal neutron experiment ($\Delta < 0.04$~meV); good fits were obtained assuming $\Delta = 0$ as shown by the solid lines.
This negligible gap energy indicates that the Se-doped Co(S$_{1-x}$Se$_{x}$)$_2$ ($x = 0.05$) compound is an excellent isotropic (or soft) ferromagnet, similar to undoped CoS$_2$~\cite{iiz75}.

\begin{figure}
 \includegraphics[scale=0.3, angle=-90]{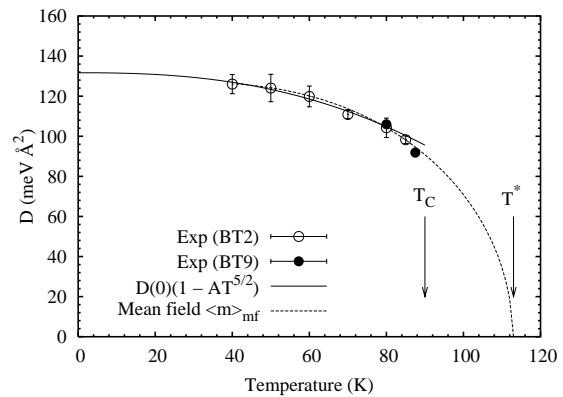}%
 \caption{Temperature dependence of the spin-wave stiffness parameter $D(T)$.
	$D(T)$ exhibits a discontinuity at the ferromagnetic transition temperature $T_{\rm C} = 90$~K, rather than decreasing smoothly to zero at $T_{\rm C}$.
	The solid curve is the fit to $D(0)(1 - AT^{5/2})$; fitted parameters were $D(0) = 131.7 \pm 2.8$~meV-\AA$^{2}$ and $A = (3.6 \pm 0.3) \times 10^{-6}$~(1/K$^{5/2}$).
	The dashed line represents the temperature dependence of the mean-field magnetization $<m>_{\rm mf}$, scaled to fit the experimentally obtained $D(T)$.
	}
	
\end{figure}

The quadratic dispersion relation enables us to express the temperature dependence of the spin-wave dispersion relation as the renormalization of the stiffness parameter $D(T)$.
Fig.~6 shows the temperature dependence of the stiffness parameter.
This can be well explained by the conventional spin-wave theory for the Heisenberg ferromagnet~\cite{dys56}, in which two-magnon interactions give rise to leading order temperature dependence in $D(T)$ as $T^{5/2}$:
\begin{equation}\label{eq2}
	D(T) = D(0)(1-AT^{5/2}).
\end{equation}
Shown by the solid line in Fig.~6 is a result of fitting to Eq.~(\ref{eq2}).
It can be seen that the two-magnon interactions can account for the temperature dependence up to $\sim 80$~K ($\sim 0.9T_{\rm C}$).
The extrapolated zero-temperature spin-wave stiffness is $D(0) = 131.7 \pm 2.8$~meV-\AA$^{2}$.

\begin{figure}
 \includegraphics[scale=0.3, angle=-90]{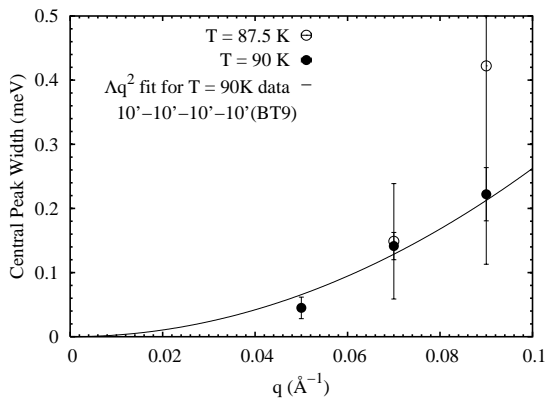}%
 \caption{$q$ dependence of the quasielastic peak width at $T = 87.5$~K $ < T_{\rm C}$ and $90$~K $ = T_{\rm C}$.
	The $q$-dependence can be explained by the diffusion law $\Gamma_{\rm q} = \Lambda q^2$ at $T = 90$~K, while the precision of the $T = 87.5$~K data is not sufficient for the fitting.}
\end{figure}

\begin{figure}
 \includegraphics[scale=0.3, angle=-90]{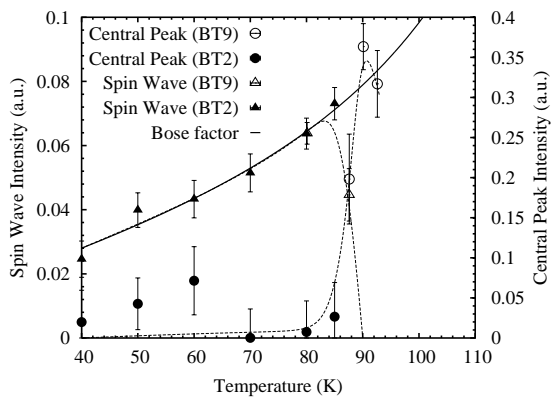}%
 \caption{Integrated intensity of the quasielastic central peak (open and closed circles) and the spin-wave peak (open and closed triangles).
	The dashed lines are guides to the eye, whereas the solid line stands for the properly-scaled Bose temperature factor $[1 - \exp(-E_q/k_{\rm B}T)]^{-1}$.
	The quasielastic component starts to develop at $T \sim 85$~K, where the spin-wave intensity shows a concomitant decrease.
	At $T = T_{\rm C}$, the spectrum is dominated by the quasielastic component, as shown in Fig.~4(b).}
\end{figure}

For conventional ferromagnets, $D(T)$ exhibits the power-law behavior in the critical temperature region, continuously decreasing to zero for $T \rightarrow T_{\rm C}$.
However, as is obvious in Fig.~6, the presently-observed $D(T)$ does not collapse at $T_{\rm C}$; $D(T)$ at $T = 87.5~{\rm K} \ (0.97\ T_{\rm C})$ is still quite large, $91.9 \pm 1.3$~meV-\AA$^{2}$.
This discontinuity in $D(T)$ at $T_{\rm C}$ is another characteristic of the first-order transition in Co(S$_{1-x}$Se$_x$)$_2$.
The discontinuity further suggests that the characteristic temperature scale $T^{*}$ for the ferromagnetic interaction strength is not $T_{\rm C}$, but rather a temperature where an extrapolation of $D(T)$ to higher temperature becomes zero.
We, thus, extrapolated $D(T)$ using the temperature dependence of the mean-field magnetization $<m>_{\rm mf}$; this extrapolation has a less theoretical rigor compared to the low-temperature extrapolation using Eq.~(\ref{eq2}), and thus the obtained $T^{*}$ should be regarded as a rough estimation.
The dashed line in Fig.~6 represents $<m>_{\rm mf}$ with $T^{*} \sim 113$~K.
We note that the obtained $T^{*}$ is reasonably close to that in undoped CoS$_2$ ($T_{\rm C} = 122$~K), suggesting that the Se doping does not significantly affect the exchange interactions.
It is known that the ratio of $D(0)$ and $k_{\rm B}T^{*}$ gives an estimate of the magnetic interaction range.
From the extrapolated $D(0)$ and $T^{*}$, we obtain $D(0)/k_{\rm B}T^{*} \sim 14$~\AA$^{2}$, which is only slightly smaller than the value ($D(0)/k_{\rm B}T_{\rm C} \sim 17$\AA$^{2}$) for the undoped CoS$_2$.
This ratio, on the other hand, is still quite large compared to conventional ferromagnetic materials~\cite{lyn95}, presumably reflecting the itinerant character of the ferromagnetism in Co(S$_{1-2}$Se$_x$)$_2$.

As mentioned earlier, the excitation spectrum shows an abrupt development of the quasielastic component as $T \rightarrow T_{\rm C}$ from below, which coexists with the well-defined spin-wave peaks (Fig.~4(b)). 
A quasielastic component may appear in the vicinity of $T_{\rm C}$ even in conventional ferromagnet due to longitudinal spin fluctuations~\cite{lyn94}.
However, the longitudinal component is generally weak and broad in energy, and appears in the temperature region where the transverse spin-waves significantly renormalize and broaden.
Consequently, it typically cannot be distinguished as a peak in unpolarized neutron experiments; indeed, polarized neutrons are necessary to detect it.
Therefore, it is unlikely to ascribe the presently-observed quasielastic component to longitudinal fluctuations.
Shown in Fig.~7 is the $q$-dependence of the quasielastic peak width $\Gamma_{\rm q}$ at $T = 87.5$ and 90~K.
Due to the large uncertainty of intensities around the elastic position, the central peak widths cannot be estimated with high precision at $T = 87.5$~K.
Nevertheless, one can definitely see increasing behavior in $\Gamma_{\rm q}$ as $q$ becomes larger (open circles). 
At $T = 90$~K, the spectrum is dominated by a single quasielastic peak (Fig.~4(a)), and thus the central-peak widths could be obtained easily.
The resulting $q$-dependence, shown by the filled circles, is reasonably approximated by the diffusion law $\Gamma_{\rm q} = \Lambda q^2$ (solid line) with an effective diffusion constant $\Lambda = 26.2 \pm 2.9$~meV-\AA$^{2}$.
This indicates that corresponding spin dynamics is of the spin-diffusion type.
It is concluded that the quasielastic peak is reminiscence of the high temperature paramagnetic fluctuations.

Fig.~8 shows the temperature dependence of the spin-wave and quasielastic peak intensity observed at $q = 0.07$~\AA$^{-1}$.
The spin-wave intensities increase monotonically as the temperature is raised to $\sim 85$~K.
This increase can be explained by the usual thermal population (or Bose) factor of the spin-wave excitations, as shown by the solid line in the figure.
However, as the quasielastic component starts to develop above $T \sim 85$~K, the spin-wave intensity shows a rapid decrease.
This suggests that the ferromagnetic to paramagnetic transition may most likely be driven by the abrupt development of the paramagnetic state, instead of the thermal excitation of the spin-waves which is the relevant mechanism for the conventional ferromagnets that exhibit a continuous, second-order transition.

\section{Discussion and Conclusions}

The ferromagnetic transition in pure CoS$_2$ has been shown to be continuous (second-order) in several experiments~\cite{hir94,hir96}.
However, the stiffness parameter $D(T)$ shows an anomalously steep temperature variation at the transition temperature~\cite{iiz75}; it continuously decreases for $T \rightarrow T_{\rm C}$ as $D(T) = D(0)(1 - T/T_{\rm C})^{\beta}$, however the exponent $\beta = 0.24 \pm 0.02$ was significantly smaller than $\beta = 0.3645\pm0.0025$ expected for three-dimensional Heisenberg ferromagnets~\cite{gui80}.
In addition, a weak but finite central component was observed in the inelastic scattering spectrum.
In the present study, we have shown that $D(T)$ does not collapse at $T_{\rm C}$ as the transition becomes clearly first-order by Se doping.
This discontinuous behavior corresponds to $\beta \rightarrow 0$ if we insist on the above power-law expression.
Furthermore, the quasielastic component appears much more prominently in the Se-doped sample.
It is concluded that the anomalously steep $D(T)$ and finite quasielastic component observed in undoped CoS$_2$ are due to the proximity to the first-order transition.

To elucidate the nature of the first-order ferromagnetic transition in Co(S$_{1-x}$Se$_x$)$_2$, it may be worthwhile to compare the present results to ferromagnetic transitions observed in the optimally doped $R$MnO$_3$ ($R$: rare-earth) class of materials, exemplified by La$_{0.7}$Ca$_{0.3}$MnO$_3$ ~\cite{lyn96,lyn00}, Nd$_{0.7}$Sr$_{0.3}$MnO$_3$ and Pr$_{0.7}$Sr$_{0.3}$MnO$_3$~\cite{fer98}.
These materials are excellent isotropic ferromagnets, but show quite different spin dynamics from conventional isotropic ferromagnets.
In particular, the ferromagnetic transition is not second-order but is discontinuous first-order, accompanied by strong relaxation effects.
The spin-wave stiffness does not collapse as $T \rightarrow T_{\rm C}$ from below, but instead a quasielastic diffusive component develops significantly in the excitation spectrum, coexisting with the well-defined spin-wave peaks at finite energy up to $T_{\rm C}$.
These anomalous features bear a striking resemblance to the present results observed in Co(S$_{1-x}$Se$_{x}$)$_2$.
In the doped $R$MnO$_3$ compounds, the coexistence of the quasielastic and spin-wave peaks is ascribed to a coexistence of ferromagnetic and paramagnetic phases~\cite{lyn96}. 
Therefore, a similar inhomogeneous mixture of the two magnetic phases may likely be realized in the Se-doped Co(S$_{1-x}$Se$_{x}$)$_2$ compound.
Indeed, such an inhomogeneous mixture was inferred in the earlier NMR experiments~\cite{pan79,yas79}.
These experiments demonstrate the existence of two distinct Co moments in the ferromagnetically ordered phase; predominant Co atoms have ordered moments of 0.9~$\mu_{\rm B}$, whereas others are non-magnetic.
The thermal variation of the bulk spontaneous magnetization for $T \rightarrow T_{\rm C}$ was mainly attributed to increasing population of the non-magnetic Co atoms, not to thermal reduction of the ordered moment.
By assuming the magnetic and non-magnetic Co atoms as ones in the ferromagnetic and paramagnetic phases, the results can be re-interpreted as the inhomogeneous mixture of the coexisting ferromagnetic and paramagnetic phases.
Moreover, phenomenological Landau theory~\cite{got97,yam01} suggests a double-well-structured free energy with minima at $<M> = 0$ and $<M> \neq 0$, giving rise to the coexisting two phases at $T_{\rm C}$.

The paramagnetic fluctuations in the doped $R$MnO$_3$ compounds are further related to formation of polarons originating from the Jahn-Teller-type lattice instability inherent in the hopping $e_g^1$ electrons~\cite{mil95,rod96,mil96,lyn97,vas99,ada00,dai00}.
As noted in the introduction, the Co$^{2+}$ ions in CoS$_2$ are in the low-spin state; $t_{2g}$ levels are fully occupied, whereas $e_g$ levels contain one electron per site.
This electronic configuration makes Co$^{2+}$ ions Jahn-Teller active.
Therefore, despite the absence of the core $t_{2g}$ magnetic moment in CoS$_2$, which makes this argument rather intuitive, we may suggest a similar lattice distortion effect in the Se-doped Co(S$_{1-x}$Se$_x$)$_2$ compound.
Neither static nor dynamic Jahn-Teller distortions have been observed to date, and further study in this direction may be intriguing.

In conclusion, we have investigated the ferromagnetic transition in Co(S$_{1-x}$Se$_x$)$_2$ ($x = 0.05$) using neutron scattering.
In the elastic experiments, we confirmed the first-order nature of the transition by the discontinuity of the ferromagnetic order parameter and SANS intensity, accompanied by the strong relaxation effects.
The inelastic scattering experiments reveal that the long-wavelength spin excitations are well explained as conventional ferromagnetic spin-waves.
The spin-wave gap was negligibly small, evidencing that the Co(S$_{1-x}$Se$_x$)$_2$ is an excellent isotropic (soft) ferromagnet.
The stiffness parameter $D(T)$ decreases modestly as the temperature is increased, however, it does not collapse as $T \rightarrow T_{\rm C}$ from below.
Instead, the prominent development of a quasielastic component was observed in the vicinity of $T_{\rm C}$, coexisting with the well-defined spin-wave peaks, whose intensities concomitantly decrease.
This suggests that the ferromagnetic to paramagnetic transition is driven by the abrupt appearance of the quasielastic fluctuations, instead of the thermal excitation of the spin-waves as for conventional ferromagnets.
A coexistence of the paramagnetic and ferromagnetic phases in the vicinity of $T_{\rm C}$ is argued.

\begin{acknowledgments}
Work at Rutgers University and University of Maryland is supported by the NSF MRSEC, DMR 00-80008.
The authors thank F.~M.~Woodward for his help in SANS data analysis.
The stay of T.J.S. at National Institute of Standards and Technology is partly supported by the Atomic Energy Division, Ministry of Education, Culture, Sports, Science and Technology, Japan.
\end{acknowledgments}



\end{document}